\documentclass{article}
\usepackage{spconf,amsmath,amssymb,cite,url}
\usepackage[utf8]{inputenc}
\usepackage{graphicx}
\usepackage{color}
\usepackage{subcaption}
\usepackage{multirow}
\usepackage{fancyhdr}

\usepackage{tikz}
\usetikzlibrary{shapes}
\usetikzlibrary{arrows.meta}
\usetikzlibrary{positioning}
\usetikzlibrary{graphs}
\usetikzlibrary{quotes}
\usetikzlibrary{fit}
\usetikzlibrary{decorations.pathreplacing}
\usetikzlibrary{calc}

\tikzset{
  state/.style={
    circle, draw=black, minimum size=2.6em,
    graphs/as={#1}
  },
  obs/.style={
    rectangle, draw=black, minimum size=2.6em,
    graphs/as={#1}
  }
}

\newcommand{\permission}{%
\begin{figure}[b]{%
\footnotesize
{\includegraphics[scale=0.7]{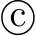}}
2019 IEEE. Personal use of this material is permitted. Permission from IEEE must be obtained for all other uses, in any current or future media, including reprinting/republishing this material for advertising or promotional purposes, creating new collective works, for resale or redistribution to servers or lists, or reuse of any copyrighted component of this work in other works.
}
\end{figure}
}

\title{Deep Polyphonic ADSR Piano Note Transcription}

\name{Rainer Kelz$^{\star}$, Sebastian B\"{o}ck$^{\star}$, Gerhard Widmer$^{\star, \dagger}$}
\address{$^{\star}$Austrian Research Institute for Artificial Intelligence, Vienna\\
  $^{\dagger}$Johannes Kepler University, Linz
}

\begin{document}
\maketitle
\thispagestyle{fancy}

\begin{abstract}
We investigate a late-fusion approach to piano transcription, combined with a strong temporal prior in the form of a handcrafted Hidden Markov Model (HMM). The network architecture under consideration is compact in terms of its number of parameters and easy to train with gradient descent. The network outputs are fused over time in the final stage to obtain note segmentations, with an HMM whose transition probabilities are chosen based on a model of attack, decay, sustain, release (ADSR) envelopes, commonly used for sound synthesis. The note segments are then subject to a final binary decision rule to reject too weak note segment hypotheses. We obtain state-of-the-art results on the MAPS dataset, and are able to outperform other approaches by a large margin, when predicting complete note regions from onsets to offsets.
\end{abstract}


\section{Introduction}\label{sec:introduction}
\permission
Polyphonic transcription is the task of extracting a symbolic score from an audio recording, regardless of how many instruments or notes are playing concurrently. For each note sounding in the recording, we would like to obtain a tuple $(s, e, n, v)$, denoting start, end, MIDI note number and optionally volume. We tackle a somewhat easier subproblem, and attempt to transcribe polyphonic recordings of a single instrument, the piano. We will ignore volume too for now, extracting only $(s, e, n)$ tuples. To this end, we pursue a multi-task deep learning approach with late fusion of the neural networks' predictions in time. Transcription of polyphonic piano music, as well as the deep learning aspect of it, is well studied in the literature \cite{boeck_2012, elowsson_2018, thome_2017, hawthorne_2018, cong_2018, sigtia_2016, kelz_2016}.

In multi-task learning, one attempts to predict multiple targets with a shared representation \cite{caruana_2012}. This can lead to improved generalization, because representations that are helpful in predicting targets for one task can be utilized to predict targets for other tasks. In our scenario, the tasks are indeed highly related, and there is much potential for representation reuse, as we train a deep convolutional neural network to simultaneously predict the onsets, intermediate note phases and offsets of piano notes. This trick of using the same groundtruth to define multiple targets at different points in time was already mentioned in \cite{caruana_2012}, chapter 8. The network architecture we use is simple, produces musically interpretable features, and has a small number of parameters. Interpreting the network outputs as emission probabilities of an HMM, combined with transition probabilities that directly encode the temporal relationships of different note phases, allows us to obtain plausible note candidates $(s, e, n)$. We can efficiently filter these candidates after decoding to discard a large amount of false positives. This combination of multi-task learning and handcrafted, causal probabilistic temporal model yields state-of-the-art performance on extracting complete notes on the widely used MAPS  piano transcription dataset \cite{emiya_2008}.

\section{Relation to previous Work}\label{sec:related_work}
It could be shown that modelling different note phases in time with different neural network outputs can be advantageous \cite{hawthorne_2018, cong_2018, elowsson_2018, caruana_2012}. The piano transcription approach in \cite{hawthorne_2018} uses two separate, bi-directional long-short term recurrent neural networks (BLSTMs) to train a pitched onset detector together with a framewise pitch detector. The onset targets and the intermediate note phase targets are supplied to separate parts of the network, with the onset predictions feeding into the BLSTMs responsible for the final note predictions. This can encourage the suppression of spurious note activities by potentially making them dependent on the presence of an onset. The authors give a few examples demonstrating such a suppression mechanism at work. However, as there are no constraints mentioned to force the desired behavior, the BLSTMs responsible for final note predictions could just as well decide to ignore the onset predictions of the second network.

In a similar vein, \cite{cong_2018, elowsson_2018} use three (or more) separate neural networks for onsets, intermediate note phases and offsets. Predictions are fused either via a handcrafted rule-based method \cite{cong_2018}, or another neural network on top \cite{elowsson_2018} to obtain symbolic notes.

We borrow this idea of using separate targets for different phases of a note, but drastically simplify the architecture to a much smaller convolutional neural network with a common representation that branches out after a few shared layers, and predicts onsets, intermediate frames, and offsets. This is in contrast to the aforementioned architectures, which neglect any potential for feature reuse, by having completely separate networks for each task (up to $6$ different networks in \cite{elowsson_2018}).

Instead of using BLSTMs or rule-based systems, we pick a different route and post-process the predictions of the network with a handcrafted HMM to obtain individual note segmentations. The states of the HMM roughly correspond to \emph{attack}, \emph{decay}, \emph{sustain} and \emph{release} (ADSR) phases of a note, along with an additional state indicating that no note is currently sounding. ADSR envelopes, as shown in \mbox{Figure \ref{fig:final_rule}a}, are commonly used in sound synthesis, governing the volume of a note from onset to offset (ADS) and a brief period afterwards (R). The HMM is not fitted using data, instead we select the transition probabilities of the model manually, and interpret the outputs of the network as emission probabilities. After the decoding step, which yields many candidate note regions, a final decision rule is subsequently applied to each note segment, taking into account the raw network outputs and discarding segments with too small activations within a segment. ADSR envelope inspired mechanisms have been used previously, for example in \cite{lee_2012} to model the temporal evolution of spectral envelopes directly.

Post-processing raw, framewise transcriptions with HMMs is a practice widely reported in the literature. The approach in \cite{vincent_2004} uses Independent Subspace Analysis to extract raw note transcription, and HMMs to model their durations.
Similarly, \cite{cazau_2017} uses two-state HMMs to post-process raw transcriptions obtained using a variety of NMF- and PLCA-based methods. In \cite{ozerov_2009}, the temporal evolution of note spectra is modeled via factorial scaled HMMs. The authors in \cite{benetos_2013} use an HMM variant that explicitly models the duration of staying in a particular state. These are only a small sample of a great variety of related approaches, as can be found in \cite{benetos_2015, ryynanen_2005, poliner_2007, emiya_2008, yakar_2013, mysore_2010}.

\section{Models}\label{sec:models}
When predicting multiple targets simultaneously with neural networks, one can consider two ends of a spectrum. One could either branch out immediately after the input layer, and thus have a separate network for each target, or one could branch out immediately before the output layers and have a shared network for all targets. We opt to use a model somewhere in the middle of this spectrum. The first few layers compute a shared representation. Based upon this representation, separate networks branch out, enabling each branch to specialize to the nature of the target it is connected to.

\subsection{Deep convolutional neural network}\label{sec:nn}
\begin{figure}
 \includegraphics[width=\columnwidth]{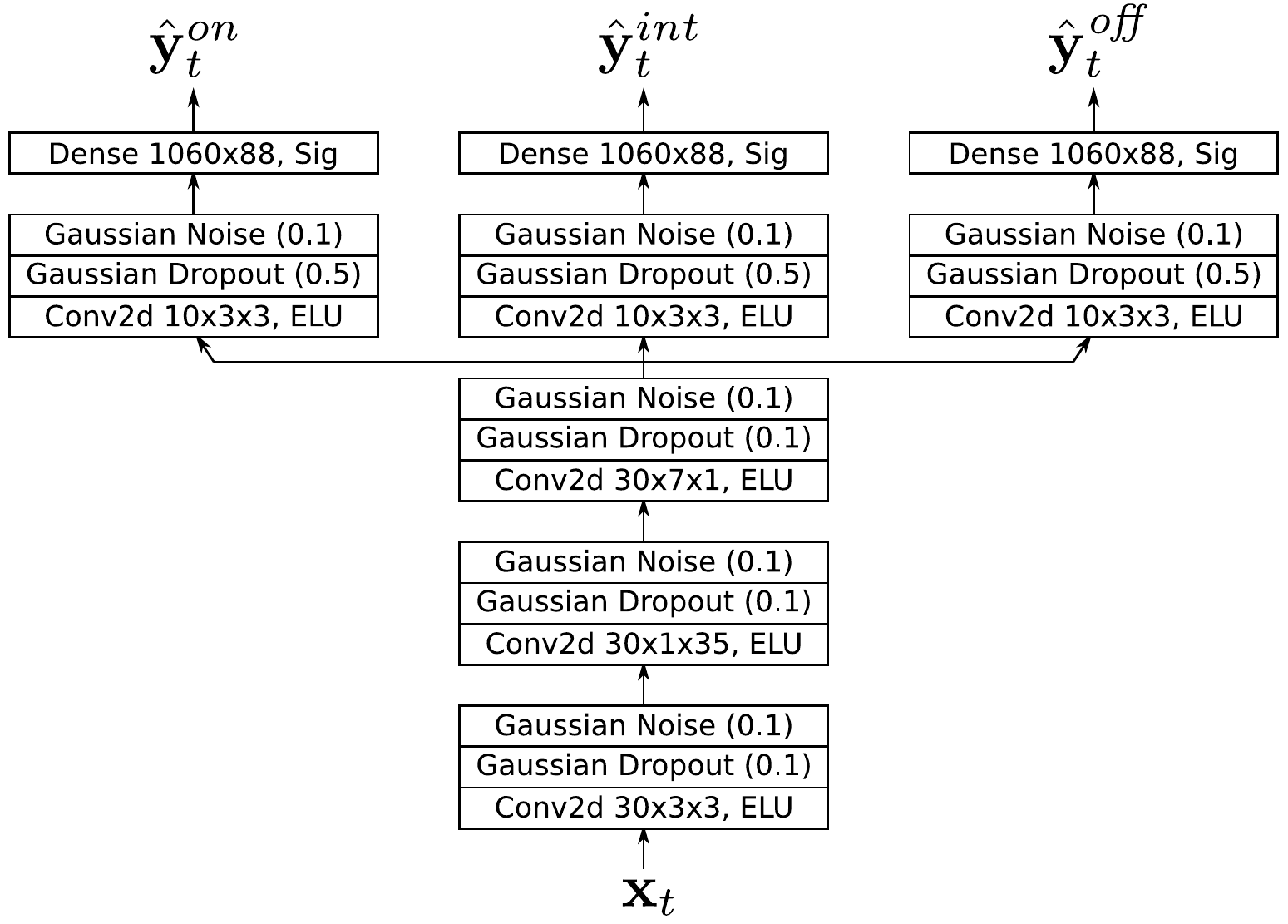}
 \caption{The architecture of our model. Arrows indicate information flow. The sizes of the convolutional kernels are given as triples $\mathrm{C} \times \mathrm{T} \times\mathrm{F}$, denoting number of channels, elongation in time and elongation in frequency dimension, respectively. The sizes of the dense layers are given as $\mathrm{I} \times \mathrm{O}$, denoting input and output dimensions, and result from concatenating all feature maps of the previous layer into a flat vector of size $\mathrm{I}$.}
 \label{fig:multiple_outputs_concept}
\end{figure}

\begin{figure}[t]
 \includegraphics[width=\columnwidth]{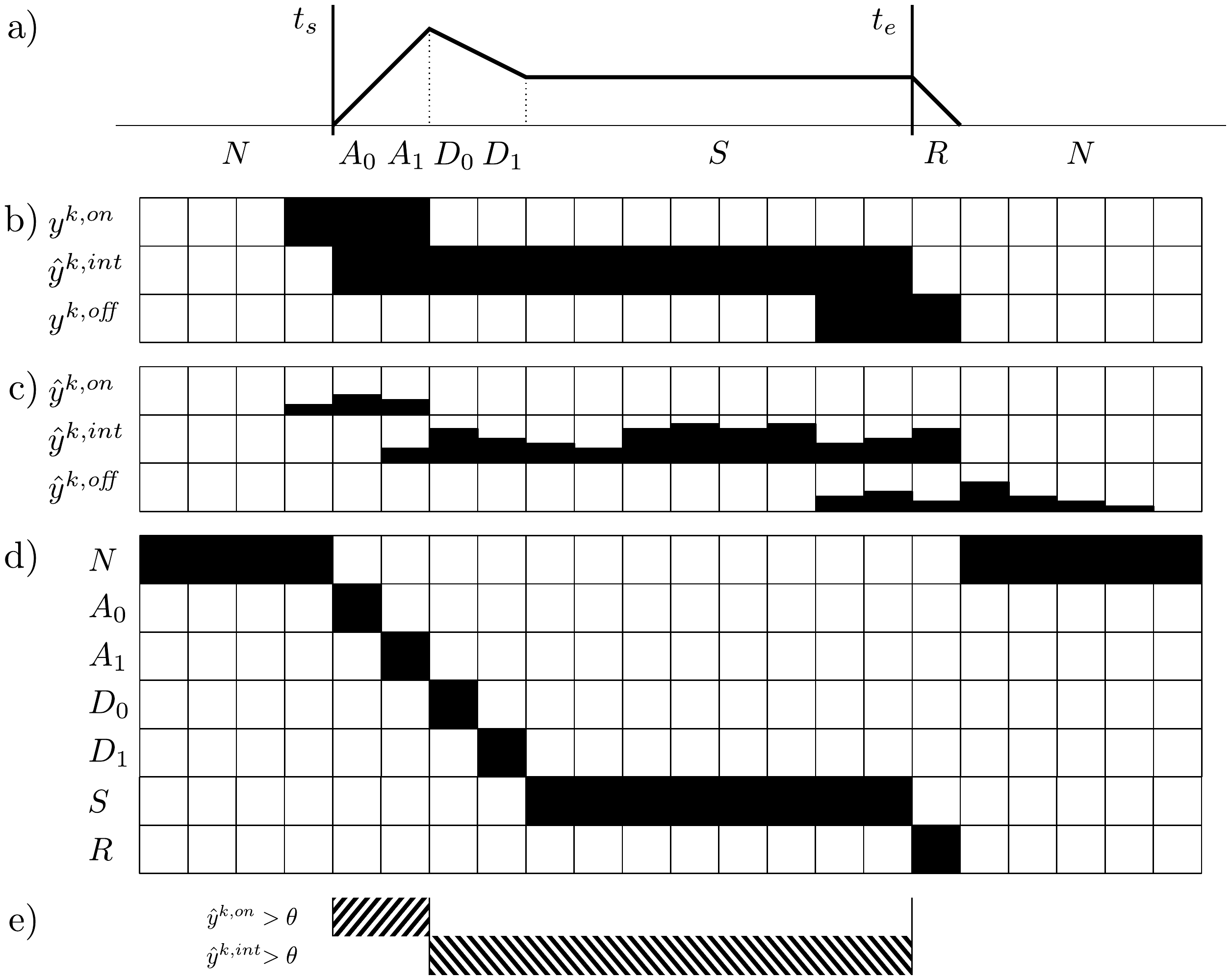}
 \caption{\textbf{a}) An idealized $ADSR$-envelope, describing the different phases of a note in time. Solid vertical lines denote the startpoint $t_s$ and endpoint $t_e$, extracted from the annotation. \textbf{b}) The targets as they are shown to the network during training. Targets with event character, such as onsets and offsets, are elongated in time to three frames. \textbf{c}) A sketch of the predictions of the network $\hat{y}^{k} \in [0, 1]$. \textbf{d}) The state space trajectory of the HMM, given the predictions. Note segmentations reach from the start of $A_0$ to the start of the $R$ state. \textbf{e}) The binary decision rule for the two different parts of the note segmentation.}
 \label{fig:final_rule}
\end{figure}

A conceptual drawing of our model architecture is depicted in Figure \ref{fig:multiple_outputs_concept}. The network input $\mathbf{x}_t \in \mathbb{R}^{c \times b}$ is a small spectrogram snippet, where $c$ denotes the number of context frames in the time dimension, and $b$ denotes the number of bins in the frequency dimension. The number $b$ is the result of passing a linear magnitude spectrogram through a filterbank with semi-logarithmically spaced, triangular filters. The filterbank has a linear response and lower resolution for the lower frequencies, and a logarithmic response for the higher frequencies. The resolution of the filtered spectrogram is approximately two bins per semitone. For all our experiments, we selected $c=11, b=144$. The temporal resolution of the model is chosen to be $50\,\mathrm{[frames/s]}$. Finally, we compute the logarithm of all magnitude bins, approximately modelling human loudness perception. The target matrix $\mathbf{y}_t \in \{0, 1\}^{88 \times 3}$ decomposes into vectors $\mathbf{y}_{t}^{on}, \mathbf{y}_{t}^{int},$ and $\mathbf{y}_{t}^{\mathit{off}}$ respectively, denoting the presence of an onset, intermediate note phase, and offset for each note in the center frame within the context window $c$. Assuming our instrument has $K$ keys, we denote the targets and predictions for the individual keys $k \in \{0 .. K - 1\}$ of the instrument at time $t$ as $\mathbf{y}_{t}^{k} \in \{0, 1 \}^{1 \times 3}$. The ground-truth annotation comes in the form of MIDI data, temporally aligned to the accompanying audio recordings. From this annotation, all three different targets are derived, as shown in Figure \ref{fig:final_rule}b, where we can also observe that for targets such as onsets and offsets, which have event character, the targets are elongated in time by one frame, to provide a denser learning signal for events.

All nonlinearities are ELUs \cite{clevert_2015}, except for sigmoid functions in the three output layers. The network is composed of small blocks with similar structure: a layer with trainable parameters, such as a convolutional or dense layer, a nonlinearity, followed by a small amount of multiplicative gaussian noise, also called ``Gaussian Dropout'' \cite{srivastava_2014}, which is sampled from $\mathcal{N}(1, (\frac{p_m}{1-p_m})^{1/2})$, followed by a small amount of additive gaussian noise, sampled from $\mathcal{N}(0, p_a)$. Please note that noise is only injected during training \footnote{Code to reproduce results is available at \url{https://github.com/rainerkelz/ICASSP19}, a pretrained model is made available in the \textit{madmom} library \cite{madmom}.}.

The related approaches in \cite{cong_2018, elowsson_2018} use three or more separate networks to obtain their predictions, which we found to be detrimental. We choose the dimensions for the convolutional kernels based on certain expectations of what features we want the network to emphasize. Kernels elongated in the time direction are supposed to emphasize loudness variations for onsets and offsets, whereas kernels elongated in frequency direction should emphasize overtone structure. The final kernel sizes settled upon can be seen in Figure \ref{fig:multiple_outputs_concept}. For an in-depth discussion on musically motivated convolutional kernel shapes see \cite{pons_2016}. The network is also compact in terms of the number of parameters, which comes down to $N = 326.394$ for the best performing model.

\subsection{Note decoding}\label{sec:hmm}
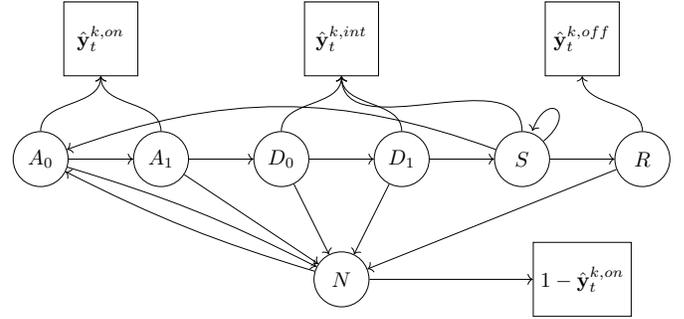
\begin{figure}
  \begin{center}
    \scalebox{0.8}{
      \begin{tikzpicture}
        \path
        (-5, 2) node[draw, circle, minimum size=2.6em] (A0) {$A_0$}
        (-3, 2) node[draw, circle, minimum size=2.6em] (A1) {$A_1$}
        
        (-1, 2) node[draw, circle, minimum size=2.6em] (D0) {$D_0$}
        (1, 2) node[draw, circle, minimum size=2.6em] (D1) {$D_1$}

        (3, 2) node[draw, circle, minimum size=2.6em] (S) {$S$}
        (5, 2) node[draw, circle, minimum size=2.6em] (R) {$R$}
        
        (0, 0) node [draw, circle, minimum size=2.6em] (N) {$N$};

        \path
        (-4, 4) node[draw, rectangle, minimum size=3.5em] (Yon) {$\hat{\mathbf{y}}_{t}^{k, on}$}
        (0, 4) node[draw, rectangle, minimum size=3.5em] (Yint) {$\hat{\mathbf{y}}_{t}^{k, int}$}
        (4, 4) node[draw, rectangle, minimum size=3.5em] (Yoff) {$\hat{\mathbf{y}}_{t}^{k, off}$}
        (4, 0) node[draw, rectangle, minimum size=3.5em] (Yonmk) {$1 - \hat{\mathbf{y}}_{t}^{k, on}$};

        \path (A0) edge [in=-90, out=90, ->] (Yon);
        \path (A1) edge [in=-90, out=90, ->] (Yon);

        \path (D0) edge [in=-90, out=90, ->] (Yint);
        \path (D1) edge [in=-90, out=90, ->] (Yint);
        \path (S) edge [in=-90, out=90, ->] (Yint);

        \path (R) edge [in=-90, out=90, ->] (Yoff);
        \path (N) edge [->] (Yonmk);
        
        \path (N) edge [bend left=5, ->] (A0);
        \path (A0) edge [bend left=5, ->] (N);

        \path (A0) edge [->] (A1);
        \path (A1) edge [->] (D0);
        \path (D0) edge [->] (D1);
        \path (D1) edge [->] (S);
        \path (S) edge [->] (R);
        \path (R) edge [->] (N);

        \path (A1) edge [->] (N);
        \path (D0) edge [->] (N);
        \path (D1) edge [->] (N);

        \path (S) edge [in=70, out=40, loop] (S);
        \path (S) edge [->, in=20, out=160] (A0);

      \end{tikzpicture}
    }
  \end{center}
 \caption{The $ADSR$-HMM model has seven states: $N$ for no note, $A_{0,1}$ for attack, $D_{0,1}$ for decay, $S$ for sustain, $R$ for release.}
 \label{fig:hmm_adsr}
\end{figure}

We will now describe the HMM-based note decoding stage for individual keys $k$. After training the network on targets $\mathbf{y}_t$, the predicted pseudo probabilities for each individual instrument key $\hat{\mathbf{y}}_t^{k}$ are interpreted as emission probabilities of an HMM. The structure of this HMM is depicted in \mbox{Figure \ref{fig:hmm_adsr}}. The transition probabilities are determined manually on the training data set, and are shared across all keys. Each key $k$ is decoded into a sequence of note segments individually, however.

%% this table has to be defined here, so it pops up on the top of the next page ... *sigh*
\begin{table*}[t]
 \begin{center}
 \begin{tabular}{|r||c|c|c||c|c|c||c|c|c|}
   \hline
   & \multicolumn{3}{c||}{Frames} & \multicolumn{3}{c||}{Note Onsets} & \multicolumn{3}{c|}{Complete Notes} \\
  Method & $\mathcal{P}$ & $\mathcal{R}$ & $\mathcal{F}$ & $\mathcal{P}$ & $\mathcal{R}$ & $\mathcal{F}$ & $\mathcal{P}$ & $\mathcal{R}$ & $\mathcal{F}$ \\
  \hline
  BLSTM \cite{hawthorne_2018} &
	%       Frames        |         Onsets        |         Notes
    % P       R       F   |   P       R       F   |   P       R       F   
  	88.53 & 70.89 & 78.30 & 84.24 & 80.67 & 82.29 & 51.32 & 49.31 & 50.22 \\
  ADSRNet &
    % notes_adsr_b (final)
    % P       R       F   |   P       R       F   |   P       R       F   
  	90.73 & 67.85 & 77.16 & 90.15 & 74.78 & 81.38 & 61.93 & 51.66 & 56.08 \\
  \hline
 \end{tabular}
\end{center}
 \caption{Experimental results on the Disklavier recordings} 
 \label{tab:maps_results}
\end{table*}

For a better understanding of the HMM structure, a sketch of an $ADSR$ envelope is provided in \mbox{Figure \ref{fig:final_rule}a}. We will call $\{A_{0,1}, D_{0,1}, S, R\}$ the sounding states, and $\{N\}$ the non-sounding state. The transition probabilities are chosen in such a way that for large pseudo probabilities for onsets, intermediate note phases and offsets, the HMM transitions through all four sounding states, as shown in \mbox{Figure \ref{fig:final_rule}d}. Smaller emission probabilities, or even larger gaps in the predictions can lead to transitions that return to the non-sounding state earlier, as we can see from the multiple arrows going into the $N$ state.
We allow for the possibility to immediately return to the $A_0$ state from $S$, starting a new note segment without going through $R$. The reason for this is, that pressing the same key rapidly in sequence leads to audio for which the network outputs only very low offset pseudo probabilities. After decoding, we keep all segments that transition at least from $A_0$ to $S$ for further processing. This effectively establishes a lower bound on the note length, given by the state sequence $\{A_{0,1}, D_{0,1}, S\}$, which yields a minimum note length of $0.1[s]$ at a framerate of $50 [\mathrm{frames/s}]$. After note segmentations have been obtained, a final rule is applied to each note segment, utilizing the raw predictions $\hat{\mathbf{y}}_{t}^{k}$, with $t \in [\mathrm{frame}(A_0), \mathrm{frame}(\mathrm{last}(S))]$, for that segment, where $\mathrm{frame}(\cdot)$ returns the frame number of an HMM state, and $\mathrm{last}(\cdot)$ returns the last state in a sequence of recurring states. If there is at least one pseudo probability $\hat{y}_t^{k,on} \geq \theta$ during the $\{A_0, A_1\}$ phases, and at least one pseudo probability $\hat{y}_t^{k,int} \geq \theta$ during the $\{D_{0,1}, S\}$ phases, the segment is kept, otherwise it is discarded. An illustration of this mechanism is shown in \mbox{Figure \ref{fig:final_rule}e}.

\section{Experiments}\label{sec:experiments}
We use the MAPS dataset \cite{emiya_2010} to train and select models. The dataset contains 210 recordings of classical piano music, rendered using 7 samplebank-based synthesizers. Additionally, there are two sets of recordings of a reproducing Disklavier piano: 30 recordings from a microphone in close proximity to the piano, and 30 recordings from a microphone farther apart, capturing additional ambient acoustic conditions, such as room reverberations.

All neural network models are trained, compared and selected \textit{only} on the synthetic sources, and finally evaluated on the Disklavier recordings, for which results are reported in Table~\ref{tab:maps_results}. Additionally, we \textit{remove} any \textit{musical overlap} from the trainset, yielding only $137$ musical recordings. We agree with \cite{hawthorne_2018} in this regard, and see this as an important step to reduce trainset bias. As the neural network model is fairly small in terms of parameters, and the training loss never reaches zero, we can assume that the model is underfitting the data to some extent. This encouraged us to hand-tune the transition probabilities of the HMM towards best performance \textit{only} on the predictions obtained from the training set, assuming the error behavior and output distribution of the neural network will be similar enough on unseen data, due to the loose fit.

Note transcription performance is determined with the same audio aligned ground-truth annotations and evaluation protocol as in \cite{hawthorne_2018}, scoring each musical piece individually and averaging performance measures across all pieces.
We utilize the \verb|mir_eval| \cite{raffel_2014} library with the following parametrization: onsets are counted as correctly transcribed, if they are within a $\pm 50 \mathrm{[ms]}$ range of the annotated onset. An offset is counted as correct if it is within $\pm 50 \mathrm{[ms]}$ or $\pm 20\%$ of the note length, whichever happens to be larger.

We can see from Table~\ref{tab:maps_results}, that the trained network outputs fairly precise predictions for all three targets, and the majority of keys. As we directly condition the start of note segments on the presence of onset predictions, this necessitates conservatism and confidence in onset and offset predictions. Even though the recall suffers for most of the measurements, due to the small size of the network, and considerable differences in acoustic conditions between train- and testset, the directly enforced constraints on what note segments should look like in time, manage to boost the recognition performance for complete notes considerably.

Results from \cite{kelz_2016, cong_2018, sigtia_2016} were omitted from Table~\ref{tab:maps_results}, because their trainsets (called ``Configuration II'') contain significant musical overlap with the testset, biasing the results.

\section{Conclusion and Future Work}
We have shown that simple, small convolutional neural networks with multiple outputs for different temporal phases of a note, together with sequential probabilistic models can achieve state-of-the-art results on a widely used piano transcription dataset.

Some potential improvements for the future include: a global model for typical note lengths, with the help of hierarchical HMMs, trying to infer fingering information from the networks' predictions, which could lead to improvements in transcribing keys which are pressed and released together.

Additionally, we would like to incorporate a \textit{post-hoc}, linear analysis of the volume a note was played at, and only then mapping it to a MIDI velocity number. We believe this to be a better model for volume, than trying to directly predict this quantity with neural networks, as done in \cite{hawthorne_2018}.

\section{Acknowledgments}
This work is supported by the European Research Council via ERC Grant Agreement 670035, project \mbox{CON ESPRESSIONE} and the Austrian Promotion Agency (FFG) under the ``BASIS, Basisprogramm'' umbrella program. The Tesla K40 used for this research was donated by the NVIDIA Corporation.

\bibliographystyle{IEEEbib}
\small{
\bibliography{master}
}
\end{document}